\documentclass[preprint, numberedappendix]{aastex} 
\usepackage{natbib}
\usepackage{graphicx}
\usepackage{amsmath,amssymb,amsfonts}
\usepackage[paper=letterpaper,margin=1in, top=1.75in]{geometry}
\usepackage{wasysym}
\usepackage{epstopdf}

\newcommand{\nogap}{\negthickspace}
\newcommand{\erfc}[1]{\mathrm{erfc} \nogap \left( #1 \right)}

\newcommand{\expf}[1]{\mathrm{exp} \left[ #1 \right]}

\newcommand{\be}{\begin{equation}}
\newcommand{\ee}{\end{equation}}
\newcommand{\bea}{\begin{eqnarray}}
\newcommand{\eea}{\end{eqnarray}}
\newcommand{\ba}{\begin{array}}
\newcommand{\ea}{\end{array}}

\newcommand{\diff}[1]{ {\mathrm{d}}{#1} }
\newcommand{\intg}[2]{ \int {#1} \; \diff{#2} }
\newcommand{\dintg}[4]{ \int_{#3}^{#4} {#1} \;\diff{#2} }
\newcommand{\delt}[1]{\delta {#1}}
\newcommand{\Delt}[1]{\Delta {#1}}

\newcommand{\pxp}[1]{ \left( #1 \right) }

\newcommand{\abs}[1]{\left| #1 \right| }

\newcommand{\usim}{\sim \nogap}
\newcommand{\mendash}{\text{\textendash}}

\newcommand{\punits}[1]{\, \mathrm{ #1 }}
\newcommand{\scinum}[2]{#1 \times 10^{#2}}

\newcommand{\degs}{ {}^{\circ} }

\newcommand{\refp}[1]{(\ref{#1})}

\newcommand{\sref}[1]{section \ref{#1}} 
\newcommand{\Sref}[1]{Section \ref{#1}}
\newcommand{\aref}[1]{appendix \ref{#1}}

\newcommand{\eqnref}[1]{equation \refp{#1}}
\newcommand{\Eqnref}[1]{Equation \refp{#1}}
\newcommand{\figref}[1]{figure \ref{#1}}
\newcommand{\Figref}[1]{Figure \ref{#1}}
\newcommand{\tabref}[1]{table \ref{#1}}

\newcommand{\commentout}[1]{}

\begin{document}

\title{Multi-messenger astronomy of gravitational-wave sources with flexible wide-area radio transient surveys}

\author{
Cregg C. Yancey\altaffilmark{1},
Brandon E. Bear\altaffilmark{2},
Bernadine Akukwe\altaffilmark{2},
Kevin Chen\altaffilmark{3},
Jayce Dowell\altaffilmark{4},
Jonathan D. Gough\altaffilmark{5},
Jonah Kanner\altaffilmark{6},
Michael Kavic\altaffilmark{7},
Kenneth Obenberger\altaffilmark{4},
Peter Shawhan\altaffilmark{1},
John H. Simonetti\altaffilmark{2},
Gregory B. Taylor\altaffilmark{4},
Jr-Wei Tsai\altaffilmark{2}
}

\altaffiltext{1}{Department of Physics, University of Maryland, College Park, MD 20742, USA}
\altaffiltext{2}{Department of Physics, Virginia Tech, Blacksburg, VA 24061, U.S.A}
\altaffiltext{3}{Department of Physics, The College of New Jersey, Ewing, NJ 08628}
\altaffiltext{4}{Department of Physics and Astronomy, University of New Mexico, Albuquerque NM, 87131.}
\altaffiltext{5}{Department of Chemistry, Lehman College,
Bronx, NY 10468}
\altaffiltext{6}{LIGO-California Institute of Technology, Pasadena, California 91125}
\altaffiltext{7}{Department of Physics, Long Island University, Brooklyn, New York 11201, U.S.A.}


\begin{abstract}

We explore opportunities
for multi-messenger astronomy using gravitational waves (GWs) and
prompt, transient low-frequency radio emission to study highly energetic
astrophysical events.  We review the literature on possible
sources of correlated emission of gravitational waves and radio
transients, highlighting proposed mechanisms that lead to a
short-duration, high-flux radio pulse originating from the merger of
two neutron stars or from a superconducting cosmic string cusp.
We discuss the detection prospects for each of these mechanisms by
low-frequency dipole array instruments such as LWA1, LOFAR and MWA.
We find that a broad range of
models may be tested by searching for radio pulses that, when de-dispersed,
are temporally and spatially coincident with a LIGO/Virgo GW trigger within a
$\usim 30$ second time window and $\usim 200 \mendash 500 \punits{deg}^{2}$ sky region.
We consider various possible observing strategies and discuss their
advantages and disadvantages.
Uniquely, for low-frequency radio arrays, dispersion can delay the
radio pulse until after low-latency GW data analysis has identified
and reported an event candidate, enabling a \emph{prompt} radio signal
to be captured by a deliberately targeted beam.
If neutron star mergers do have detectable prompt radio emissions, a
coincident search with the GW detector network and low-frequency
radio arrays could increase the LIGO/Virgo effective search volume by
up to a factor of $\usim 2$.
For some models, we also map the parameter space that may be
constrained by non-detections.
\end{abstract}



\pagebreak
\newpage

\section{INTRODUCTION}

\label{sec:introduction}

Throughout the history of modern astronomy, transient emissions
captured by a wide range of instruments have revealed a fascinating
variety of energetic astrophysical events.
For instance, the discovery of gamma-ray bursts (GRBs) beginning in the late 1960s \citep{RWKlebesadel1973} challenged astronomers to explain the origin of remarkable high-energy transients with rapid variability, some as short as a fraction of a second.  Further detections with better directional information, provided by the BATSE instrument on the Compton Gamma Ray Observatory, established that GRBs are extragalactic \citep{BATSE1992}, but multi-wavelength observations were the key to further characterizing GRBs (e.g., with the watershed GRBs 970228 and 970508 detected and localized by the BeppoSAX satellite) and identifying some of the objects that produce them, beginning with GRB 980425 = SN 1998bw \citep{Kulkarni1998}.

Transient astronomy is now established in all bands of the electromagnetic spectrum, with different strategies depending on instrumental capabilities and the accessible population of sources.  A full review of transient surveys is beyond the scope of this paper, but we point out that wide-field instruments have a natural advantage.  For instance, the Fermi Gamma-ray Burst Monitor \citep{FermiGBM} views more than half of the sky at any time, while current optical survey efforts such as iPTF \citep{PTF2009}, CRTS \citep{CRTS2009}, MASTER \citep{MASTER2004}, and Dark Energy Survey \citep{DECam2015} feature optical imagers with fields of view of several square degrees, systematically visiting large areas of the sky.
Fast radio bursts (FRBs)---isolated, short, highly dispersed radio pulses---are presently a hot topic, with several intriguing events reported \citep{Lorimer:2007qn,DThornton2013, EPetroff2015, SBurke-Spolaor2014, LGSpitler2014} and numerous theories as to their origin\footnote{Aside from the ``perytons'' now attributed to a microwave oven \citep{Perytons}.}, but no clear picture yet.  Distinct from single pulses from Galactic neutron stars \citep{MAMcLaughlin2006}, radio flares from shocks \citep{SRKulkarni}, and late-peaking radio afterglows \citep{ENakar2011,Ghirlanda2014,Metzger2015}, FRBs may be another prompt signature of familiar phenomena such as supernovae or GRBs, or else a hallmark of something more exotic.  It might turn out that most FRBs are produced by nearby flaring stars \citep{arXiv150701002}, but here we assume that at least some FRBs are produced by compact objects which also emit gravitational waves.

All FRBs reported to date have been detected in the 1.4 GHz band.  However, it has been argued \citep{Lorimer2013,MWAProspects} that FRBs should also be detectable at lower frequencies by relatively new facilities such as the Long Wavelength Array \citep[LWA,][]{EllingsonLWA1}, the LOw Frequency ARray \citep[LOFAR,][]{LOFAR2013} and Murchison Widefield Array \citep[MWA,][]{Bowman2013}.  These facilities consist of clusters of hundreds of dipole antennas with back-end electronics and digital processing that combine the antenna signals with configurable phase offsets.  They are \emph{flexible}, capable of forming instantaneously steerable beams or of operating in a wide-area mode, with selectable central frequency and bandwidth; specific capabilities depend on the back-end processing.  However, no FRBs have been detected so far in searches performed with these instruments \citep{Coenen2014}.

The advent of sensitive gravitational-wave observatories, namely LIGO \citep{BPAbbott2009}, Virgo \citep{Accadia:2012zzb}, and GEO600 \citep{Grote:2010zz} provides an additional means of observing the transient sky through gravitational waves (GWs) and may reveal the physical engine driving the transients.  Gravitational waves can provide direct information regarding the masses and motions associated with an observed transient, as this information is encoded in the gravitational wave's waveform \citep{JFaber2012,KKiuchi2009,MMaggiore2008}.  Such information is not readily attainable from electromagnetic emissions, which generally arise from reprocessed energy or outflows and are subject to absorption and scattering.  In contrast, GWs penetrate even dense environments without modification, but have, so far, remained elusive to detection.  Coupling GW observations with another independent astrophysical messenger, such as radio transients, could significantly improve the sensitivity of detection for gravitational waves.  Thus, it is beneficial to combine electromagnetic and gravitational wave observations to study the internal dynamics driving high-energy astrophysical transients \citep{BloomWP}.

In this paper, we consider \emph{multi-messenger} astronomy enabled by coincidence of gravitational waves and prompt low-frequency radio emissions (pulses) to study short-duration (up to $\usim 1 \punits{sec}$), high-energy transients.  Two specific reasons motivate this approach.  The first is that there are several common sources for correlated emission of gravitational waves and low-frequency radio.  The second reason is that both the GW and radio instruments are capable of observing large areas of the sky; in fact, the GW detectors respond to waves arriving from all directions, guaranteeing overlap.  Prior consideration has been given to an effort such as this \citep{VPredio2010}, however, the confirmation of FRBs and the availability of better instruments makes such an effort of even greater interest.  And, as we will discuss, rapid coordination now being put in place makes it possible for low-frequency radio instruments to point in the direction of a GW event candidate in time to catch the dispersion-delayed prompt pulse.

Below, \sref{sec:instruments} summarizes the properties of the instruments which are relevant for the discussion in the rest of the paper.  \Sref{sec:sources} outlines various mechanisms for binary neutron star mergers and superconducting cosmic strings as sources for correlated emission of gravitational waves and radio transients.  \Sref{sec:observation_method} details the coincidence method and derives an appropriate coincidence time window, while \sref{sec:observation_strategies} discusses three different observational strategies available for multi-messenger astronomy with gravitational waves and radio transients.
\Sref{sec:sensitivity_improvements} calculates estimated improvements in detection sensitivity as a result of joint observations.  Finally, \sref{sec:summary} summarizes the important points discussed throughout the paper and mentions potential astrophysics that may result from this effort.


\section{INSTRUMENTS}

\label{sec:instruments}

\subsection{Gravitational Wave Detectors}

Modern GW detectors use laser interferometry to detect tiny variations in the local spacetime metric due to a passing gravitational wave, specifically by measuring differential changes in the lengths of two orthogonal arms using optical cavities and feedback to measure length changes as small as $\usim 10^{-19} \punits{m}$.  For the tensor wave polarizations predicted by the general theory of relativity, the detectors act as quadrupolar antennas, responding to incoming waves from all directions (even through the earth) with just a few discrete null directions.

Several gravitational waves detectors are available for detection of gravitational waves, and several more are expected to be commissioned for use in the near future.
After a long-planned major upgrade, the Advanced LIGO \citep[aLIGO,][]{aLIGO} detectors are now operational in Hanford, Washington and Livingston, Louisiana, with preliminary sensitivites already more than three times better than the original LIGO detectors.  Their sensitivities are expected to improve by a further factor of $\usim 3$ over the next few years \citep{JAasi2013}.
Each detector responds to gravitational waves with frequencies in the range $\usim 10 \mendash 5000 \punits{Hz}$, with best sensitivity around $100 \punits{Hz}$.  The detectors are expected to detect binary neutron star inspirals out to a distance of $\usim 450 \punits{Mpc}$ for optimal sky location and orientation of the binary, or $\usim 200 \punits{Mpc}$ averaged over all directions and orientations.
The GEO600 detector has been an important testbed for the development of advanced technologies \citep{GEO600adv} and has collected data for many years, but does not have comparable sensitivity to aLIGO, so it will not be discussed further in this paper.
The Advanced Virgo \citep[AdVirgo,][]{AdVirgo} detector, located in Cascina near Pisa, Italy, has a similar design to aLIGO but with arms $3 \punits{km}$ long.  It is expected to be operational by 2017 and to ultimately reach a sensitivity about 2/3 that of aLIGO.  aLIGO and AdVirgo will operate as a coherent network, sensing the same GW signals and sharing data for joint analysis.  A new 3-km-long detector, KAGRA \citep{KAGRA}, is currently under construction in Japan and will join the network later this decade.  An additional aLIGO detector is planned for installation at a new observatory in India early next decade.  Each additional detector enhances the detection, direction determination, and parameter estimation capabilities of the network.

\subsection{Low-frequency Radio Facilities}

The first completed LWA station, LWA1 \citep{GBTaylor2012}, is a phased-array radio telescope composed of 258 dipole-antenna pairs which is co-located with the VLA in New Mexico.  It is sensitive to radio frequencies in the range $10 \mendash 88 \punits{MHz}$.  The signal processing system is capable of forming 4 independently steerable beams.  Each beam has a full-width at half maximum (FWHM) at zenith of $2.2 \degs \times \left( 74 \punits{MHz} / \nu \right) \sec^2(Z)$, where $\nu$ is the frequency and $Z$ is the zenith angle \citep{LWAtechnical}.  Assuming a typical zenith angle of $30 \degs$, this gives a FWHM of $5.7 \degs$ at $38 \punits{MHz}$, which corresponds to an area of $\usim 26 \punits{deg}^{2}$ for each beam.  In addition to synthesized beams, LWA1 supports two all-sky modes, transient-buffer narrow (TBN) and transient-buffer wide (TBW), wherein it coherently captures and records data from all its nodes.  The TBN all-sky mode allows continuous data recording
with a $70 \punits{kHz}$ bandwidth.  The TBW all-sky mode allows data recording at the full $78 \punits{MHz}$ bandwidth supported by LWA1, but recording can only occur in $61 \punits{ms}$ bursts at 5 minute intervals.  Coupled to the LWA1 is the Prototype All-Sky Imager (PASI), which is a software correlation and imaging back-end that generates all-sky images from LWA1's TBN-mode data \citep{Obenberger2015}.

In addition to LWA1, which has been operating for the past 4 years, new LWA stations are coming on-line
at Owen's Valley, California (LWA-OVRO) and at Sevilleta, New Mexico (LWA-SV).  Both of these instruments
will provide the capability to survey the entire visible hemisphere at much broader bandwidths than LWA1, 
typically 10 MHz or more compared to just 70 kHz.

LOFAR \citep{LOFAR2013} is a phased-array radio interferometer composed of dipole antenna stations located in the Netherlands and across Europe.  It is designed to be sensitive to the low-frequency range from $10 \mendash 240 \punits{MHz}$ with a large field of view (FoV).  There are currently $18$ stations in the Netherlands in an area known as the LOFAR core, $5$ in Germany, and the UK, France, and Sweden each have $1$.  Each core station is comprised of 96 low-band antennas (LBAs) and two sub-stations, each with 24 high-band antenna (HBAs) tiles.  The LBAs are designed to operate between $10 \mendash 90 \punits{MHz}$, and the HBAs measure between $110 \mendash 240 \punits{MHz}$.  Similar to LWA1, LOFAR is capable of observing in all-sky mode and producing steerable beams, with a FoV for the central part of 6 stations' (``Superterp'') tied-array beams of $\sim 5'$.


The MWA \citep{Bowman2013} is an array of $2048$ dual-polarization dipole antennas located in the Shire of Murchison in Western Australia, optimized for $80 \mendash 300 \punits{MHz}$.  The antennas are arranged within $128$ tiles as $4\times4$ dipole arrays, and each tile is capable of beam-forming an electronically steerable beam with a field of view of $25^\circ$ at $150 \punits{MHz}$.  
Thus each tile sees 625 square degrees.  
We will focus on LWA1 and LOFAR in the discussion 
below because of the larger instantaneous fields-of-view, greater overlap on the sky, and other advantages 
offered by the lowest frequencies.


\section{SOURCES}

\label{sec:sources}

\subsection{Neutron Star Binary Mergers}

\label{sec:ns-ns_mergers}
Compact binary mergers are expected to be a primary source of GWs observable by ground-based gravitational-wave detectors.  Such binaries would include neutron star-neutron star (NS-NS) binaries, neutron star-black hole (NS-BH) binaries, and black hole-black hole (BH-BH) binaries.  Of primary concern here are NS-NS binaries, which are expected to be a prime common source for radio and GW transients; further, these systems are the leading candidate as the progenitor of short, hard-spectrum GRBs \citep{JFaber2012} \citep{KKiuchi2010}.

\subsubsection{NS-NS Mergers and Gravitational Wave Emission}
\label{sec:ns-ns_GW_production}

NS-NS binaries form from the stellar evolution of binary star systems containing $8 \mendash 10 \punits{M_{\astrosun}}$ stars \citep{JFaber2012}.  As the NS-NS binary evolves, its orbit decays due to GW emission.  Eventually, the individual neutron stars merge.  The lifetime of the NS-NS binary can be divided into three phases \citep{JFaber2012}:
\begin{itemize}
\item Inspiral: NSs undergo a decaying orbit due to GW emission.  This is the longest phase of the NS-NS binary's lifetime.
\item Merger: NSs fall directly toward one another and collide.  Merger begins in the last $1 \mendash 2$ orbits at the end of the inspiral phase.
\item Ring-down: immediately after merger, the final remnant may oscillate or spin, emitting lower amplitude GWs if its mass distribution is non-axisymmetric.
\end{itemize}
The merger and ring-down phases occupy only the last $10 \mendash 30 \punits{ms}$ of the binary's lifespan before resulting in the formation of a black hole.  In some cases, an intermediary hyper-massive neutron star (HMNS), $2.7 \mendash 2.9 \punits{M_{\astrosun}}$ supported by thermal pressure and rotation, can form and exist for $5 \mendash 25 \punits{ms}$ after merger \citep{TWBaumgarte2000, KHotokezaka2011, KKiuchi2009, YSekiguchi2011}, before finally collapsing to a black hole.  Transient radio emission can occur either just prior to the merger or after the merger in the ring-down phase or during the collapse to a black hole \citep{CPalenzuela2008}. In particular, the formation of a HMNS is important to the emission model of Pshirkov and Postnov \citep{MSPshirkov2010}, discussed later.

The GWs are emitted throughout the inspiral phase with increasing frequency and amplitude (the characteristic ``chirp'' signature), with a peak burst of GWs occurring during the merger.  GW emission after merger and during the ring-down phase can vary considerably depending on the binary mass, the equation of state for supernuclear-density matter, and the formation of the HMNS \citep{KHotokezaka2013}.

\citet{JAbadie2010} review predictions for the rates of compact binary mergers, and the expected detection rates for various GW detectors.  The best constraints are on the NS-NS merger results, which are extrapolated from the known observed binary pulsars in our Galaxy.  For aLIGO, they say that a likely detection rate would be about 40 events per year, at aLIGO design sensitivity, with a possible range of 0.4 to 400 per year.  Current plans for aLIGO include a gradual increase in sensitivity, with predicted 
range limits for GW observable NS-NS mergers to start at about 60~Mpc in 2015, then become 100-140~Mpc in 2016-2018, and achieve 200-215~Mpc in 2019-2020  \citep{JAasi2013}.  In this scenario the expected event rate for aLIGO is between 0.08 and 8 NS-NS mergers per year in 2015.  

In addition to NS-NS mergers, a merger of a neutron star with a black hole (NS-BH) would
also produce detectable GWs, and could also produce a radio frequency transient.  
For example, simulations by \citet{VPaschalidis2013} suggest the potential for NS-BH mergers to produce precursor radio signals in the kHz range.  While the discussion
that follows highlights NS-NS mergers, some of the described models for radio emission could apply equally well to NS-BH mergers.  

\subsubsection{NS-NS Radio Transient Production Mechanisms}
\label{sec:ns-ns_radio_transient_production}

Models of radio emission exist for different epochs centered around the moment of merger.  Below we review mechanisms for producing \emph{prompt} radio emission, within seconds before or after the time of the NS-NS merger.  A radio afterglow signal is also expected, due to jetted outflow interacting with the ambient interstellar medium \citep{ENakar2011,Ghirlanda2014,Metzger2015}. Follow-up observations of afterglow emission from NS-NS merger events would be of great interest. However, this type of long timescale emission does not allow for de-dispersion of the radio signal. Thus it is not possible to establish the type of direct temporal link between a GW burst and a prompt radio transient that is considered here.  Therefore this kind of emission is not a target for the type of coincident search described in this paper.



\paragraph{Pre-merger: The Model of Hansen and Lyutikov}
\label{sec:pre-merger_Hanson_Lyutikov}


The mechanism studied by \citet{BMSHansen2001} results in the emission of coherent, low-frequency radiation in the few seconds prior to the NS-NS merger.
In this model one NS would be a recycled pulsar, spinning relatively rapidly (spin period $\sim$1--100~ms), with a relatively low magnetic field strength ($B_r \sim10^{9}$--$10^{11}$~G).  The other NS has a relatively higher magnetic field strength (and thus may be referred to as a ``magnetar'', $B_m \sim10^{12}$--$10^{15}$~G) and has spun down to a low spin period ($\sim$10--1000~s).  Stronger $B_m$ values result in stronger transient pulses before the merger.


The interaction of the recycled pulsar with the external magnetic field of the magnetar leads to an extraction of energy from the pulsar's spin and orbital motion.  Hansen and Lyutikov model this situation using a perfectly conducting sphere (representing the pulsar) moving in a uniform magnetic field $\mathbf{B_0}$ with velocity $v$ and spin angular velocity $\Omega$.  The sphere will exclude the external magnetic field from its interior, producing an induced dipole field of its own.  The resultant total magnetic field is given by
\be
\mathbf{B}_{tot}=\mathbf{B_0}+\frac{R^3}{2r^3}\mathbf{B_0}-\frac{3R^3(\mathbf{B_0}\cdot\mathbf{r})\mathbf{r}}{2r^5}	
\ee
where $R$ is the radius of the sphere, and $\mathbf{r}$ is the displacement vector from the center of the sphere (with magnitude $r$).

The orbital motion and spin of the pulsar induces a charge density on its surface.  The electric field produced by the charge density will accelerate charges to relativistic energies $\gamma m_e c^2$ in an attempt to cancel the component of the electric field parallel to the total magnetic field. This produces a primary beam of electrons, and, if they are accelerated to a sufficient energy ($\gamma \sim 10^6$), the primaries produce curvature photons and secondary electron-positron pairs, in much the same way as the case of a classical pulsar but with the important difference that there are no closed field lines; so, energy extraction occurs over the entire pulsar's surface and not just at the polar caps.

The orbital and spin energy extracted from the pulsar (mainly from the primary beam of particles) is expected to be
\be
L \sim 4 \pi R^2 n_{GL} \gamma_{max} m_e c^3 \sim 3.1 \times 10^{36} \, {\rm erg \, s^{-1}},
\label{Lbeam}
\ee
where $n_{GL}$ is a typical beam density; $n_{GL} \sim \Omega B_0/2\pi e c$ for acceleration of charges induced by rotation, and $n_{GL} \sim vB_0/e c R$ for acceleration of charges induced by orbital motion.   Assuming an efficiency of $\epsilon \sim 0.1$ for the conversion of this energy into radio emission (arguing from the classical pulsar case), Hansen and Lyutikov estimate the observable flux density at 400~MHz to be
\be
F_{\nu} \sim 2.1 \, {\rm mJy}\, \left(\frac{\nu}{400\,{\rm MHz}}\right)^{-2} \, \frac{\epsilon}{0.1} \left(\frac{D}{100\,{\rm Mpc}}\right)^{-2} B_{15}^{2/3} \, a_7^{-5/2} \label{Radio},
\ee
where $D$ is the distance to the binary, $B_{15}$ is the magnetar field strength in units of 10$^{15} \punits{G}$, and $a_7$ is the distance of the pulsar from the magnetar in units of $10^7 \punits{cm}$.  Scaling to lower frequency $\nu$ we will parametrize as $\propto \nu^{-\alpha}$, where $\alpha\sim2$, arguing from observations of a typical pulsar.   Note that a turnover of this power law at some low frequency is inevitable, but we have ignored this issue here.

Hansen and Lyutikov conclude the emission is weak and would not be easily detectable by current instruments for radio transient searches.  
However, from the discussion of the LWA1 sensitivity in Appendix (\ref{sec:sensitivity_LWA}), we find we can detect events that Hansen and Lyutikov describe to distances of
30~Mpc at 38~MHz and 20~Mpc at 74~MHz, for detections near the zenith, for an emitted pulse of 10~s.  The distance limit drops to about 10~Mpc at 38~MHz for zenith angles of about 50$^\circ$ due to sky noise correlation across the array (as described in Appendix (\ref{sec:sensitivity_LWA})).  LOFAR can detect events to comparable distances of 20~Mpc.  Temporal broadening of the pulse by the combined effects of dispersion (across a frequency channel) and interstellar/intergalactic scattering will negligibly broaden the emitted pulse before its arrival at the telescope.  These distances are comparable to the 60~Mpc predicted distance limit for the 2015 aLIGO system.

It is currently understood that approximately 1\% or less of NS-NS binary systems consist of a magnetar \citep{SBPopov2006}.  Therefore, considering typical LWA1 beam widths and an average NS-NS coalescence rate of $1 \punits{Mpc}^{-3} \punits{Myr}^{-1}$ \citep{JAbadie2010}, the detection rate for this mechanism is of order $10^{-6}$ per year at $38 \punits{MHz}$ and $10^{-7}$ per year at $74 \punits{MHz}$, for one beam along the zenith.

\paragraph{During Merger: The Model of Pshirkov and Postnov}
\label{sec:during_merger_Pshirkov_Postnov}


\citet{MSPshirkov2010} consider a model in which radio emission
occurs just after the merger, but prior to the collapse of the resulting object to a black hole.  Low-energy, pulsar-like emissions result from energy transfer from the differential rotation of the merger remnant into the surrounding magnetic field. 
The emitted pulse is expected to have a temporal length of order 10~ms, i.e., the time period between the formation of the merged object, and its subsequent collapse to a black hole.

This model takes the total energy pumped into the magnetic field from the differential rotation energy as  
\begin{equation}
B^2 R^3 \sim (\Delta\Omega/\Omega)^2 \Delta E,
\end{equation}
in cgs units, where $B$ is the magnetic field, $R$ is the characteristic radius of the region occupied by the field ($R\approx10^6$~cm, here), $\Delta\Omega/\Omega$ is the factor characterizing the differential rotation, and $\Delta E$ is the full rotational energy (expected to be the same as the orbital energy at the merger, $\Delta E \sim 10^{53}$~erg).  Thus the magnetic field could be increased to as large as $10^{16}$~G by the merger, but they take $B\sim10^{15}$~G, to be conservative.  Then, as in the standard discussion of a pulsar, the rotating magnetic dipole radiates an electromagnetic luminosity of
\begin{equation}
\dot{E} \sim \frac{\Omega^4 B^2 R^6}{c^3}.
\end{equation}
For $B\sim10^{15}$~G, $R\sim10^6$~cm, and $\Omega\sim 6000$~s$^{-1}$ (the orbital value at merger), they find $\dot{E}\sim10^{50}$~erg~s$^{-1}$.

Pshirkov and Postnov treat the problem phenomenologically, assuming a fraction $\eta$ of this energy loss rate is output as radio emission, and adopting a value of $\eta$ which is weakly dependent on $\dot{E}$, thus nominally $\eta =10^{-5} (\dot{E}/10^{35}\ {\rm erg\ s}^{-1})^\gamma$, with $1/2<\gamma<0$ as suggested by observations of rapidly rotating pulsars.  Therefore, taking $\gamma=0$ (the most optimistic scenario), they find the flux density that would be observed \textit{ignoring temporal scatter-broadening of the pulse} is \begin{equation}
F\sim 8000\ \dot{E}_{50} D_{\rm Gpc}^{-2} \ \ {\rm Jy}
\end{equation}
at an observing frequency of $\nu=100$~MHz,
where $\dot{E}_{50}=\dot{E}/(10^{50}\,\rm erg\,s^{-1})$.
Following Pshirkov and Postnov in assuming a spectral index of $-2$, and that the scatter-broadened width of an observed pulse will be $\Delta t = 100\, D_{\rm Gpc}^2\, \nu_{120}^{-4}$~seconds, where $D_{\rm Gpc}$ is the distance to the source in units of Gpc, and $\nu_{120}$ is the observing frequency in units of 120~MHz, we obtain a final observed flux density of 
\begin{equation}
f_\nu \sim 0.6\ \dot{E}_{50} D_{\rm Gpc}^{-4} \nu_{120}^2  \ \ {\rm Jy}
\label{PP2010emission}
\end{equation}
in their most optimistic scenario.

If
we use the full discussion of Appendix (\ref{sec:sensitivity_LWA}), including an appropriate model of scattering, we can provide a more realistic assessment of this model.  We also fully characterize their expected radio luminosity including the range of efficiency values and values for $\gamma$ they discuss, as
\begin{equation}
L_{\rm radio} = 10^\delta \left(   \frac{\dot{E}}{10^{35}\ {\rm erg\ s}^{-1}} \right)^\gamma  \dot{E}.
\label{PPparameters}
\end{equation}
Following Pshirkov and Postnov in setting $\delta=-5$ and $\gamma=0$ gives detection distances for LWA1 and LOFAR to be 2.7~Gpc and 3.7~Gpc, respectively.
For a large fraction of the phase space of the parameters $\delta$ and $\gamma$ these limiting distances are larger than those obtainable by even the final version of aLIGO.  The 10~ms emitted pulse will have broadened to about 0.44~s at 38~MHz, during propagation.

\begin{figure}[htp]
\centering
\includegraphics[width=.4\textwidth]{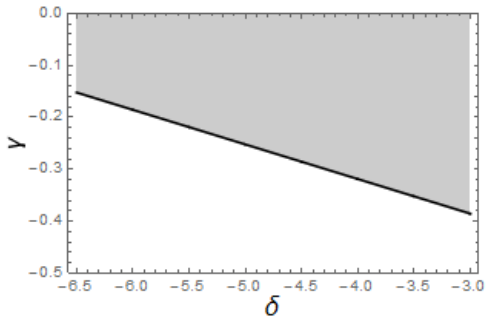}
\caption{The excluded section of the parameter space in the Pshirkov \& Postnov model at $38$~MHz with the signal-to-noise ratio threshold set to $10$ is shown. The shaded region represents the excluded portion following no signal detections in a coincidence search of aLIGO and LWA1.  The solid curve corresponds to $200$~Mpc, the projected average detection distance for aLIGO.}
\label{fig:PPparameterspace}
\end{figure}

Finally, there is potential to reconsider the applicability of the phenomenological parameter space of $\delta$ and $\gamma$ considered by Pshirkov and Postnov.  As seen in \figref{fig:PPparameterspace}, a large portion of the parameter space will be constrained if no detections are observed. 

This mechanism is highly favorable for a radio-GW coincident search for two reasons.  It is expected that the transient radio pulse and the GW are emitted approximately simultaneously.  As well, simulations suggest that the magnetic field amplification would proceed in a manner similar to that expected by this mechanism \citep{2006Sci...312..719P, LRezzolla2011, JZrake2013}.  Thus such amplification is considered to occur in most cases of NS-NS mergers, rather than requiring one NS to be a magnetar.  Using the best case scenario discussed above, the detection rate for LWA1 is $\sim 10$ detections per year at 74~MHz and $\sim 100$ detections per year at 38~MHz for one beam pointing along the zenith.  These detection rates follow closely with the expected detection rates of aLIGO.  However, the potential for the amplification of the magnetic field depends on minimal disruption from GW emission from the merged object.  Thus, a balance of conditions for maximal magnetic field amplification and GW emission is required in a coincidence search as described in this paper, and not all cases will provide the optimal scenario discussed here.

\paragraph{GW Induced MHD Emission of Radio: The Model of Moortgat and Kuijpers}

\label{sec:GW_induced_MHD_emission}

A very intriguing source of correlated GW and radio emission, 
which is emblematic of the effort proposed here, is radio transient production directly from propagation of a strong 
gravitational wave through a magnetohydrodynamic (MHD) plasma, as discussed by \citet{JMoortgat2004}.  The GWs emitted in an exothermic astrophysical process can excite magnetohydrodynamic waves as they propagate through a plasma.  The electromagnetic radiation production is primarily caused by inverse Compton radiation 
modulated at the frequency of the gravitational wave and a Lorentz factor of the particles in the plasma jet. The radiation has a frequency in the radio, as does the jet, and only escapes the jet when the frequency $f=\omega_p/\sqrt{\gamma_s}$, where $\omega_p$ is the non-relativistic plasma frequency of particles in the observer frame, and $\gamma_s$ is the secondary particles Lorentz factor. It was demonstrated in \citet{JMoortgat2004} that this process would result in the emission of coherent radiation which would be detectable in radio transient arrays.
\commentout{
The anticipated duration of the electromagnetic pulse in such a scenario is similar to that of Gamma-Ray Burst (GRB) models, $\Delta t=\frac{L}{c\gamma_s^2}$  It has thus been proposed that a GRB detection can be utilized in the identification of a gravitational wave \citep{Usov}. Theoretically, a binary star merger scenario would produce a flux at 30MHz of duration 3min \citep{MHD}.
}

Using the calculation in Appendix (\ref{sec:sensitivity_LWA}) we obtain distance limits for detection of this source by LWA1 and LOFAR to be several orders of magnitude larger than those obtained by aLIGO. \commentout{about 12~Gpc at 38~MHz and 16~Gpc at 74~MHz, and by LOFAR to be about 24~Gpc} \commentout{larger than the observable universe, considerably larger than any limiting distance obtained by aLIGO}  Using a similar method to the previously discussed radio emission mechanisms, the detection rate for LWA1 is $\sim 10^{3}$--$10^{4}$ per year for both frequencies along the zenith.  These extremely large detection rates suggest that, in the absence of positive detections, LWA1 and LOFAR will strongly constrain the parameter space of this model.

\subsection{Superconducting Cosmic Strings}

\label{sec:cosmic_strings}

Cosmic strings are one-dimensional topological defects which are postulated to have formed during the early symmetry-breaking phase transitions of the universe.  Their existence conforms with predictions made by various models for elementary particles~\citep{Vilenkin&Shellard:1994}, and their activity is thought to be related to several observable astrophysical phenomena.

The length ${l}$, energy $\xi$, and lifetime $\tau$ of a typical cosmic string loop at cosmological time ${t}$ are given by
\begin{align}
&l \sim \alpha {t} \\
&\xi \sim \mu {l} \sim \mu \left( \alpha {t} \right) \\
&\tau \sim \left( \frac{\alpha}{\Gamma_g {G} \mu} \right) {t} \sim {t},
\end{align}
where $\alpha$ is a dimensionless length parameter, $\mu$ is the string tension, $\Gamma_g$ is a calculated numerical constant equal to $\sim 50$ \citep{Vilenkin&Shellard:1994}, and $G$ is the gravitational constant. In the expression above and the rest of this subsection we take $\hbar=1=c$. Although its exact value is not well known, $\alpha$ can be approximated by assuming a relation to the gravitational back-reaction~\citep{1988cost.book.....B}, such that
\begin{equation}
\alpha \sim \Gamma_g {G} \mu.
\end{equation}
Cosmic strings are suspected to be sensitive to external electromagnetic fields, becoming superconducting current carriers when moving through magnetic cosmic backgrounds~\citep{Witten85}.
Superconducting cosmic strings are fluid, current-bearing loops which oscillate under their own tension $\mu$, given by
\begin{equation}
\mu \sim \eta^2,
\end{equation}
where $\eta$ is the symmetry breaking scale of the string.  String currents may have originated from the application of an external electric field $E$ found in cosmic settings.  A superconducting loop with charge carrier of charge ${e}$ in a magnetic field $B$ generates an alternating current of amplitude~\citep{Vilenkin87}
\begin{equation}
{i_0} \sim 0.1 {e^2} {B} {l}.
\end{equation}
Because electromagnetic emission from strings may contribute to detectable distortions of the cosmic microwave background spectrum, constraints on ${G} \mu$ of possible strings exist, so that ${G} \mu < 6.1 \times 10^{-7}$ \citep{Pogosian:2008am}.  

Of particular interest for the type of correlated search discussed here is the concurrent emission of detectable, $\mathcal{O}(100\,{\rm Hz})$, GWs~\citep{TDamour2000, TDamour2005} and low-frequency electromagnetic waves~\citep{2011PhRvD..84h5006B, Vachaspati:2008} from oscillations along the strings. 
The value of ${G} \mu$, which characterizes the gravitational interactions of strings, has been shown to correlate to detectable wave emissions for values down to ${G} \mu \sim 10^{-13}$~\citep{TDamour2000}.  Although strings of ${G} \mu << 10^{-7}$ do not typically emit recognizable signals~\citep{TDamour2005}, when a string undergoes a cusp event, the magnitude of GW emission is temporarily amplified.

A cusp event can be described as a naturally occurring solution to the equations of motion of a cosmic string loop in which a point on the oscillating loop reaches near luminal velocity for a short period of time~\citep{Vilenkin87}.  Because large scale cosmic strings behave classically and these solutions arise naturally, cusp events should create repetitive bursts of detectable gravitational radiation~\citep{TDamour2000}. LIGO is capable of effectively searching for GW bursts from cosmic string cusp events using matched filtering because the waveform for such bursts is well understood \citep{PhysRevLett.85.3761}. A search for cusp events conducted with LIGO was used to constrain the string tension to be $G\mu<10^{-8}$ \citep{PhysRevLett.112.131101}.

Although energy radiated by such a string is predominantly gravitational~\citep{VBerezinsky2001} at cusp events, the current may be heightened to a terminal value
\begin{equation}
{i_{max}} \sim {e} \eta.
\end{equation}
The enhanced current at a cusp allows for the relativistic beaming of a powerful pulse of low-frequency electromagnetic radiation.
In fact, it was claimed by \citep{Vachaspati:2008} that the fast radio burst observed by \citep{Lorimer:2007qn} could have been produced by a superconducting cosmic cusp. They derived the fluence of such a burst to be
\begin{equation}
F \sim b i_0^2 \frac{l^2}{d^2}
           e^{-a \omega l \theta^3} \ \ \ {\rm if} \ \ \ 
a \omega l \theta^3 > 1
\end{equation}
where $a$ and $b$ are constants that depend on the shape of 
the cusp with nominal values of $a \sim 1$, $b \sim 1$, $\omega$ is the angular frequency of the cusp, $\theta$ is the angle from the beam direction and $d$ is the distance to the cosmic string. 
Making use of the observed properties of the Lorimer pulse this becomes
\begin{eqnarray}
F_{\rm obs} 
&\approx& 10^{-23} e^{-4 \nu/\nu_0} ~ {\rm \frac{ergs}{cm^2\,Hz}}
\label{obsF}
\end{eqnarray}
where $\nu_0=1.4$ GHz and $\nu$ is the observing frequency.

Following the analysis of Vachaspati and using the known sensitivity of LWA1, \cite{mikeInPrep} determined that the event rate from cusp events of superconducting cosmic strings could be as high as $\sim1$ per day for that instrument. The absence of positive detections at this relatively high event rate would allow for strong constraints to be set on the allowed parameter space of superconducting cosmic string models. 
Given that a superconducting cosmic string is suspected to emit both gravitational and electromagnetic radiation in detectable ranges during cusp events under the same parameters, performing coincident observations of both spectra of radiation would provide a unique means for the discovery and study of superconducting cosmic strings.


\section{COINCIDENCE ANALYSIS DETAILS}
\label{sec:observation_method}

  

The search algorithms for GW transients and radio transients each work by processing their respective data streams and identifying significant event candidates, or ``triggers''.  Each trigger is characterized by an arrival time, a strength (measured by a detection statistic),
and directional information that comes either from analysis of multi-sensor data or from the pointing direction of a synthesized beam that recorded the signal.
The goal of coincidence analysis is to determine whether these trigger properties are consistent with being from the same astrophysical source at some position in the sky.

Gravitational-wave triggers are obtained from analysis of detector output using various methods to determine the existence of signals consistent with a gravitational wave passing through the network of detectors.  To search for burst-type signals, so-called ``coherent" methods that rely on cross-correlations between detector data \citep{SKlimenko2005, SKlimenko2011} are used.  To search for inspiral or cosmic string cusp signals, matched filtering is used in conjunction with time coincidence and source parameter consistency between the detectors in the network \citep{JAbadie2012, BAllen2005}.  The coalescence time of a binary merger or the central time of a short burst or cosmic string cusp can be determined with a precision of order $1 \punits{ms}$.  The sky position of the source is determined only probabilistically, and rather poorly due to the long wavelength and low amplitude of detectable signals.  A ``sky-map'' is calculated for each event candidate, containing the probability density as a function of position, and the probably regions typically have areas of a few hundred square degrees.

Radio triggers are generated by identifying signals above a given threshold in a de-dispersed time series. Such a search is carried out over many DMs across the full bandwidth. The observed DM and the central observing frequency can be used to determine the dispersive delay of the pulse, as discussed below.

The following sections detail the coincidence conditions that are suitable for these triggers.  Similar considerations have been discussed previously for joint surveys between GWs and neutrinos  \citep{bartos2012} and between GWs and GRBs \citep{dietz2013}.

\subsection{Temporal Coincidence}
\label{sec:temporal_coincidence}

To identify coincident events in GW and radio detectors, we must understand the measurement uncertainties in the trigger times, as well as the possible intrinsic time offset between GW and radio emissions, so that we can use an appropriate time window.  This
requires accounting for the delay in propagation of radio signals through the interstellar medium (ISM)
and the relative timing of emission, which has to be estimated from theoretical models.

\begin{figure}[htp]
\centering
\includegraphics[width=.4\textwidth]{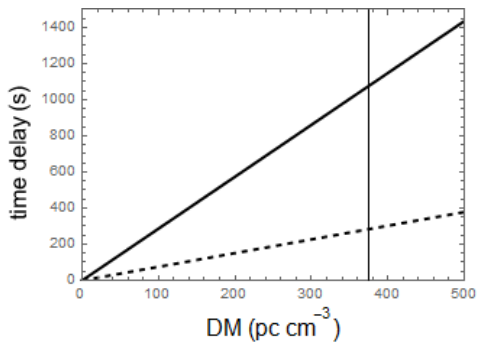}
\caption{Temporal delay of radio signals due to dispersion at $38$~MHz (solid) and $74$~MHz (dashed).  The vertical line represents the dispersion measure of the FRB commonly known as the Lorimer burst, measured at DM=$375 \punits{pc} \punits{cm}^{-3}$ \citep{Lorimer:2007qn}.}
\label{fig:DM_time_delay}
\end{figure}

Electromagnetic signals propagating through the ISM are delayed due to dispersion by an amount
\begin{equation}
\Delt{t_{\mathrm{disp}}}=(777.9 \, \mathrm{s}) \pxp{\frac{\mathrm{DM}}{300 \punits{pc} \punits{cm}^{-3}}} \pxp{\frac{40 \punits{MHz}}{\nu}}^{2}, \label{eqn:dispersion_delay}
\end{equation}
where $\nu$ is the electromagnetic frequency in megahertz and DM is the \emph{dispersion measure} given by
\begin{equation}
\mathrm{DM} = \intg{n_{e}(\vec{x})}{x} \label{eqn:dispersion_measure},
\end{equation}
where $n_{e}(\vec{x})$ is the free electron density of the medium through which the signal propagates.
For reported FRBs and extragalactic sources, typical dispersion measures can be in the range $\usim 300 \mendash 10^{3} \punits{pc} \punits{cm}^{-3}$ \citep{JMCordes2002, DRLorimer2013}.  The choice of fiducial values in eq.\ (\ref{eqn:dispersion_delay}) make it clear that dispersion delays can be many minutes for low-frequency radio observations of extragalactic sources (\figref{fig:DM_time_delay}).  However, using the measured DM, it is straightforward to infer when the radio signal would have arrived at the receiver if there had been no dispersion delay.  This is referred to as the de-dispersed time and is calculated by modifying the radio trigger time $t_O$ by the dispersion delay for the reference frequency at which the trigger was reported:
\begin{equation}
t_{\mathrm{de-disp}} = t_{O} - \Delt{t_{\mathrm{disp}}} \,.
\label{eqn:de-dispersed_time}
\end{equation}

The dispersion delay uncertainty, which will contribute to the temporal coincidence time window, has the following components, combined in quadrature
\begin{align}
\delt{t_{\mathrm{DM}}} &= (777.9 \, \mathrm{s}) \pxp{\frac{ \delt{(\mathrm{DM})} }{300 \punits{pc} \punits{cm}^{-3}}} \pxp{\frac{40 \punits{MHz}}{\nu}}^{2} \\
\delt{t_{\mathrm{freq}}} &= 2 \cdot (777.9 \, \mathrm{s}) \pxp{\frac{\mathrm{DM}}{300 \punits{pc} \punits{cm}^{-3}}} \pxp{\frac{40 \punits{MHz}}{\nu}}^{3} \cdot \pxp{\frac{\delt{\nu}}{ 40 \punits{MHz} }},
\end{align}
where $\delt{(\mathrm{DM})}$ is the dispersion-measure uncertainty and $\delt{ \nu }$ is the frequency bandwidth
of each discrete channel in the processed radio data.  The uncertainty in the radio observation time is a quadrature combination of the intrinsic pulse-width, the channel time-width (due to partitioning the radio signal into discrete, finite bandwidth channels), and the pulse-width broadening due to scattering (see \aref{sec:sensitivity_LWA}).
The channel time-width and pulse-width broadening are given by
\begin{equation}
\delt{t_{\mathrm{chan}}} = ( \scinum{2.5}{-8} \, \mathrm{s} ) \pxp{\frac{40 \punits{MHz}}{\delt{\nu}}}
\label{eqn:channel_error}
\end{equation}
and
\begin{equation}
\tau_{\mathrm{scatt}} \approx (\scinum{2}{-2} \punits{s}) \cdot \mathrm{D}_{\mathrm{Gpc}}^{-1/5} \cdot \nu_{\mathrm{40}}^{-3.9},
\label{eqn:scatter_broadening}
\end{equation}
respectively, where $\mathrm{D}_{\mathrm{Gpc}}$ is the distance to the source in gigaparsecs and $\nu_{\mathrm{40}}$ is the radio frequency in units of $40 \punits{MHz}$.  Intrinsic pulse-widths are expected to be in the range $\usim 0.01 \mendash 1 \punits{sec}$.

Typical analysis parameters for LWA1 data (summarized in \tabref{tab:LWA_error_params}) are $\nu = 38 \punits{MHz}$, $\delt{ \nu } = \scinum{7.32}{-3} \punits{MHz}$, and $\delt{(\mathrm{DM})} = 0.002 \cdot \mathrm{DM}$.
\begin{table}[ptb]
\begin{center}
\begin{tabular}{c c}
Quantity  &  Value \\ 
\hline \hline
$\nu$  &  $38 \punits{MHz}$\\
$\delt{ \nu }$  &  $\scinum{7.32}{-3} \punits{MHz}$ \\
$\delt{ \mathrm{DM} }$  &  $\usim 0.6 \mendash 2 \punits{pc} \punits{cm}^{-3}$ \\
\hline
\end{tabular}
\end{center}
\caption{\label{tab:LWA_error_params} Typical LWA1 instrument parameters that contribute to error in the de-dispersed time.}
\end{table}
For extragalactic sources observable by both aLIGO/AdVirgo and LWA1, distances can range from $8 \punits{kpc}$ (distance from Earth to the closest satellite dwarf galaxy of the Milky Way) to $\usim 400 \punits{Mpc}$ (the sensitivity limit of aLIGO for NS-NS mergers).  Thus, limits in the various contributions to the uncertainty in the de-dispersed time can be obtained and are summarized in \tabref{tab:de-dispersed_time_error_contribs}.
\begin{table}[ptb]
\begin{center}
\begin{tabular}{c c c}
Term & Minimum ($\punits{s}$) & Maximum ($\punits{s}$) \\
\hline \hline
$\delt{t_{\mathrm{DM}}}$ & 1.7 & 5.7 \\
$\delt{t_{\mathrm{freq}}}$ & 0.33 & 1.1 \\
$\delt{t_{\mathrm{pulse}}}$ & 0.010 & 1.0 \\
$\tau_{scatt}$ & 0.03 & 0.26 \\
$\delt{t_{\mathrm{chan}}}$ & 0.00014 & 0.00014 \\
\hline
\end{tabular}
\end{center}
\caption{\label{tab:de-dispersed_time_error_contribs} Calculated ranges of contributions to the de-dispersed time uncertainty.}
\end{table}
The total uncertainty in the de-dispersed time can be found by adding the individual sources of uncertainty in quadrature, which gives a range of
\begin{equation}
1.7 \punits{s} \lesssim \delt{ t_{\mathrm{de-disp}} } \lesssim 5.9 \punits{s}.
\label{eqn:de-dispersed_time_error_range}
\end{equation}

The relative timing of emission at the source creates an offset in the arrival time of the gravitational wave relative to the de-dispersed time.  As a convention, positive values in the timing correspond to GW emission after the radio transient (GW arrives after the de-dispersed time) and negative values correspond to GW emission before the radio transient (GW arrives before the de-dispersed time).  The relative timing is estimated from the models mentioned in \sref{sec:sources} to be in the range of $-35 \punits{ms}$ to $+10 \punits{s}$.  Combining this offset with a $2 \sigma$ uncertainty window in the de-dispersed time, an asymmetric temporal coincidence condition is obtained:
\begin{equation}
-11.8 \punits{s} \lesssim t_{\mathrm{GW}} - t_{\mathrm{de-disp}} \lesssim 21.8 \punits{s}
\label{eqn:temporal_coinc}
\end{equation}
where $t_{\mathrm{GW}}$ is the time of the GW trigger.

\subsection{Spatial Coincidence}
\label{sec:spatial_coincidence}

Each GW trigger has an associated source position reconstruction, i.e. a sky-map, that provides the probability of a source being at a particular location in the sky \citep{SKlimenko2011}.  For the spatial coincidence, the radio beam is compared with this sky-map for overlap within the $90 \%$ confidence region.  Typical 90\% confidence areas of GW sky-maps over the next few years are expected to be $\usim 500 \punits{deg}^2$ for the two LIGO detectors, and $\usim 200 \punits{deg}^2$ for the network of three detectors including Virgo \citep{Singer2014, Berry2015}, improving as more GW detectors are added \citep{JAasi2013}.
It is conceivable to weight the overlap using the radio beam's power pattern function \citep{EllingsonLWA1} to account for edge cases, similar to what is proposed by \citet{bartos2012} except that low-frequency radio beam sizes are much larger than high-energy neutrino directional errors.  Events for which there is no overlap between the radio beam and the $90 \%$ confidence region are discarded.

\section{OBSERVING STRATEGIES}
\label{sec:observation_strategies}

Dipole array radio antennas have the versatility that they can be operated
in either a directed ``beamed'' configuration, using aperture synthesis to collect 
wide bandwidth data over selected, relatively narrow sky regions, or in a lower resolution ``all-sky'' mode that sweeps a large overhead area of the sky \citep{Kocz2015, Ellingson2013}.  This allows for several joint observation strategies, which we consider here.  These strategies are similar to other joint observation efforts involving gravitational waves and other observable counterparts \citep{loocup_methods, Singer2014, bartos2012, nissanke2012}.  In all these observation strategies, the coincidence method from \sref{sec:observation_method} is used to determine the coincidence of observed gravitational waves and radio transients.

\subsection{All-Sky Joint Survey}
\label{all_sky_joint_survey}

Ideally, a joint survey would be limited only by geometry. The GW detector network responds to GW signals arriving from all directions, though with some direction-dependent antenna factors. Avoiding the horizon, a radio array should in principle be able to observe the sky above to zenith angles of perhaps $60 \degs$, representing 25\% coverage of the entire celestial sphere. The temporal and spatial coincidence conditions described in section 4 would apply, using whatever spatial resolution is achieved by the radio data analysis.

In practice, current wide-area searches have technical limitations from back-end and/or signal processing architecture which reduce the sensitivity and spatial resolution that would ideally be achievable. For example, the LWA1 PASI system images the entire sky to $60 \degs$ zenith angle continuously, but, with 75 kHz of bandwidth, it has an order of magnitude less sensitivity than an LWA1 beam \citep{Obenberger2015}. Also with such a narrow bandwidth, PASI has no ability to measure the DM of a transient or to calculate the dispersion delay. So even if radio emission from a GW source turned out to be very strong, PASI would have limited ability to characterize the radio signal or distinguish it from terrestrial interference.

However it is already known that these sources are not extremely bright given the fact that within 13,000 hours of data\footnote{8400 hours centered at 38 MHz, 1900 at 52 MHz, 1400 at 74 MHz, and 1300 at various frequencies between 10 and 88 MHz} PASI has detected no convincing astronomical transients occurring on 5 s timescales \citep{Obenberger2015}. Therefore if EM counterparts are to be found the sensitivity needs to be enhanced above that of PASI. LWA-OVRO and LWA-SV have much larger bandwidths than PASI, significantly increasing the sensitivity. Despite this, like PASI they still use long integration times of $\sim$ 5 s compared to beams (~50 nano s), which decreases the sensitivity to any pulses shorter than this. 

All-sky imagers typically do not track across the sky; they simply phase to zenith (zero delay), correlate, and image. This is usually adequate if the integrations are short. However, in order to perform de-dispersion, images need to be stacked from a large range of times. For instance, to search images from 40 to 50 MHz at a DM of 200 pc cm$^{-3}$, would require stacking images from as far back as 3 minutes (for the lowest frequency), and up to 15 minutes at a DM of 1000 pc cm$^{-3}$. On these timescales the sources in the sky move enough that smearing would occur and sensitivity would decrease. Therefore some method of tracking would need to be implemented to prevent this. 

One method would be to perform multiple correlations at a set of phase centers around zenith for each time stamp. The phase centers would be chosen such that they had the same Declination (Dec) but varying right ascensions (RA). For each DM, images could be selected such that they would be centered at the same RA and Dec across all frequencies, despite the time differences. This method would keep the angular distribution of all sources constant and preserve the portion of sky represented in each pixel throughout the full frequency range used. The stacked images could then simply be searched through image subtraction, source removal, or other source finding methods. This method would be computationally intensive and may be unfeasible for the backends of LWA-OVRO or LWA-SV.

A less computationally intensive method would be to only correlate and image once for each frequency and time and stack pixels of given RAs and DECs. However due to projection effects, the amount of sky represented in each pixel would change as the sky rotates, and each frequency would have a different amount of sky represented in the pixels for a given RA and DEC. This would inherently decrease the SNR of any DM, getting worse at higher DMs, and lower frequencies. 

Given these challenges it would still be worthwhile to pursue the all-sky approach, given the extreme field-of-view and large positional error of GW detectors. Furthermore this approach may result in the discovery of other dispersed pulses unrelated to GW sources.

Aside from the LWA telescopes, other wide-area radio transient surveys may cover large areas but with modest instantaneous fields of view \citep[e.g.,][]{Coenen2014}, so that their chance of capturing the counterpart of a random GW event is correspondingly reduced.

\subsection{GW Triggered Observation}
\label{GW_triggered_observations}

The large dispersion delay for low-frequency radio pulses (eq.\ \ref{eqn:dispersion_delay}) creates an exciting possibility to initiate radio observations in response to GW trigger alerts received from the LIGO/Virgo network.  In this case, one or more synthesized beams can be pointed at the sky region(s) associated with the GW trigger \textit{before} the radio pulse arrives, allowing observation of any \text{prompt} radio emission from the source.  Assuming that an alert can be generated and communicated, and the radio facility can respond rapidly enough in a target-of-opportunity mode, the chance of success should be similar to that of an all-sky joint survey.

For a radio observation frequency of $38 \punits{MHz}$ and DMs in the range $200 \mendash 10^{3} \punits{pc} \punits{cm}^{-3}$, delays can be anywhere from 10 to 45 minutes.  These times are comparable to the latency associated with reporting GW triggers in the last run of the initial LIGO/Virgo network \citep{loocup_methods}, and there is an effort underway to release aLIGO/AdVirgo triggers even faster \citep{ShawhanRapidAlerts} to a network of partner astronomers.  Teams from LWA1, LOFAR and MWA are among the groups preparing to receive and act on GW triggers.\footnote{https://gw-astronomy.org/wiki/LV\_EM/PublicParticipatingGroups}

\begin{figure}[htp]
\begin{center}
\includegraphics[width=0.6\textwidth]{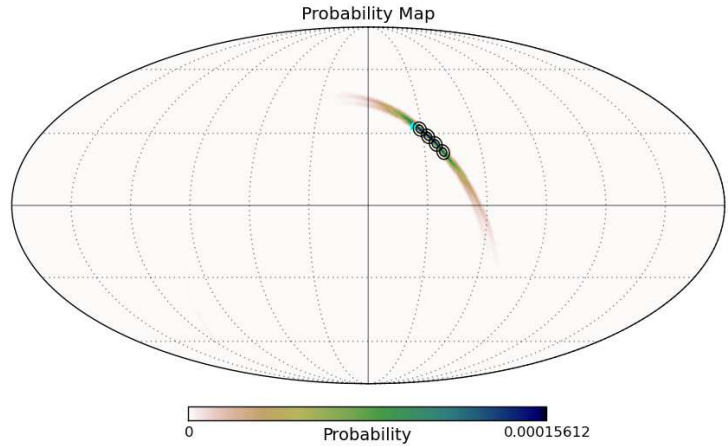}
\end{center}
\caption{\label{fig:map} A representative probability map of a candidate event during the first two years of aLIGO and AdVirgo observations \citep{2014ApJ...795..105S} with LWA1 beams superimposed. The color gradient shows the probability per square degree. Such a map will be sent as part of each GW trigger alert. LWA1 can form 4 beams simultaneous which can be used to tile the probability map as shown. Each beam is represented by two circles, one for each tuning. In this case the outer circle is the estimated beam size at 25.85 MHz and the inner circle is for 45.45 MHz.}
\end{figure}

LWA1 has recently developed two automated systems for responding to such triggers, the Heuristic Automation for LWA1 system (HAL) and the Burst Early Response Triggering system (BERT). The development of these two systems can be leveraged to conduct this triggered search.  HAL in particular is able to respond to triggers in as little as two minutes.  The LWA1 beam size and capability of forming up to four simultaneous beams allow it to cover a significant fraction of a typical GW trigger sky region (assuming it is above the horizon), as illustrated in \figref{fig:map}.
Determining spatial coincidence is not necessary in this case, as it is automatic.  Because pointed radio beams are being used, the full $19.6 \punits{MHz}$ bandwidth is available to measure the DM of a pulse and to study the radio spectrum structure of the source emission, opening the possibility to discriminate radio emission mechanisms.

GW triggered observations have the advantage that nearly all GW events reported from the LIGO/Virgo network can be tested for radio counterparts.  However, it has the disadvantage that it requires rapid coordination, and a population of radio events below a particular DM (probably around $200$) will be missed
due to the latency in issuing GW trigger alerts combined with latencies in the response of radio observatories to those alerts.


\subsection{Beamed-Radio Joint Survey}
\label{beamed_radio_joint_survey}

If a radio facility is not able to respond rapidly to alerts and re-point beams, it can still carry out a systematic survey using beams.  Radio transient and GW triggers can then be
tested for coincidence in accordance with the methods described in \sref{sec:observation_method}.
This approach has the advantage that data analysis can proceed offline, without the need for rapid communication or target-of-opportunity scheduling.
It also is suitable for prompt radio counterparts to GW events that have small DMs such the dispersion delay is too short to enable GW-triggered observations.
Using beamed-mode observation makes full use of the radio array's bandwidth, allowing DMs to be determined fairly precisely, which is critical to the temporal coincidence.
Additionally, for any radio transient that is found, the timing and directional information from the radio observations can be used as constraints for a deeper search of archived GW data, similar to what is done for GRBs \citep{JAbadie2010b}.

The major disadvantage of this approach is that instantaneous coverage of the sky is limited by the size and number of the radio beams, so that capturing an event would require a great deal of luck.  For instance, even at the lowest usable frequencies, four LWA1 beams cover no more than $\sim 1 \%$ of the sky.
As it is expected that only $\usim 40$ NS-NS mergers occur per year within the aLIGO sensitivity volume, a joint detection using this mode would be very rare, even taking advantage of the sensitivity improvements (discussed in \sref{sec:sensitivity_improvements}) that comes with a coincident search.

\commentout{
\subsection{Candidate Discrimination}
\label{sec:candidate_discrimination}
Coincidences are characterized by a figure-of-merit function that combines the gravitational-wave detection statistic $\rho$ with the radio signal SNR, and rate histograms of the resultant figure-of-merit values are generated.  To estimate the rate of background noise coincidence, the observation times for the gravitational-wave triggers and radio transient triggers are offset by a series of time shifts larger than the coincidence time-window.  Poisson statistics of background coincidences is used to obtain a discriminator threshold for the figure-of-merit values, denoted as the figure-of-merit cut-off, with the choice that candidate coincidences (i.e. those coincidences likely to correspond to a real event) have at least a $3 \sigma$ significance above the background.  This corresponds to the figure-of-merit value whose background noise rate is no greater than $\Lambda \sim \scinum{2.7}{-3} \punits{yr}^{-1}$, denoted as the false-alarm rate threshold; this threshold would be different for a different choice in candidate significance.  Coincidences whose figure-of-merit is above the figure-of-merit cut-off are considered candidate coincidences.

} 

\section{SENSITIVITY IMPROVEMENTS}
\label{sec:sensitivity_improvements}

Assuming that joint emission does occur, a coincidence analysis with GWs and radio transients makes it possible to detect somewhat weaker signals than using either observation individually \emph{if} the event occurs at a sky position which is visible to both instruments.  Since the GW network has roughly omnidirectional sensitivity, the coincidence region is limited by the radio facility.  For instance, a radio array instrument capable of pointing to a zenith angle of $60^\circ$ views a fraction $f_\mathrm{sky} = 1/4$ of the full celestial sphere.  Source detection is enhanced, then, within that fraction of the sky.

For any analysis, the significance of an event candidate can be quantified in terms of the false-alarm rate due to instrumental noise fluctuations and accidental coincidences of unrelated transients.  For a coincidence analysis with simple thresholds on the GW and radio trigger samples, the joint false-alarm rate can be estimated as
\begin{equation}
R_{\mathrm{joint}} = f_\mathrm{sky} \, R_{\mathrm{GW}} \, R_{\mathrm{radio}} \, t_{\mathrm{w}} f_c \, ,
\label{eqn:joint_false-alarm_rate}
\end{equation}
where $R_{\mathrm{GW}}$ is the full-sky gravitational-wave trigger rate and $R_{\mathrm{radio}}$ is the radio trigger rate in some mode with chosen detection thresholds.  $t_{\mathrm{w}}$ is the coincidence time window obtained from \eqnref{eqn:temporal_coinc},
$\usim 33.6 \punits{s}$,
while the factor $f_c$ represents the additional selective power of a spatial coincidence requirement (and potentially other coincidence criteria).

All gravitational-wave data collected to date contains non-gaussian instrument noise that dominates the GW trigger rate in the weak- and moderate-signal regime.  The distribution of instrumental triggers (``background'') is estimated by offsetting the data streams from the different GW detectors by a series of time shifts larger than the coincidence time window
and re-running the coherent analysis.  This re-samples the effect of non-gaussian noise while suppressing the possible contribution from astrophysical signals.  Using past LIGO data as a guide, the gravitational-wave trigger rate may be parametrized using the heuristic given in \citet{JAasi2013} that an increase in the GW detection statistic threshold $\rho$ by 1 unit corresponds to a factor $\usim 100$ reduction in the gravitational-wave trigger rate, and $\rho = 12$ corresponds to a trigger rate of $\usim 10^{-2} \punits{yr}^{-1}$, which is the nominal requirement to have high confidence in an event candidate.  This yields a functional form for the full-sky trigger rate:
\begin{equation}
R_{\mathrm{GW}}(\rho) \approx 100^{(11 - \rho)} \punits{yr}^{-1} \,.
\label{eqn:gw_false-alarm_rate}
\end{equation}
The actual sky region, radio trigger rate and spatial coincidence factor $f_c$ depend on the search strategy being followed.  Here we discuss each case:

For a wide-area radio transient search, $f_\mathrm{sky}=0.25$, but an apparent signal identified somewhere in the searched sky area will have only a chance of overlapping the sky-map of an unrelated GW trigger occurring at a consistent time, so $f_c < 1$.  For a typical GW sky-map area of $\usim 400 \punits{deg}^2$ within the quarter of the sky visible to the radio array, $f_c \approx 0.04$.  The radio trigger rate will depend critically on the signal-to-noise-ratio (SNR) threshold used in the wide area search, and the population of real transients.  For instance, if the radio search is tuned to produce an average of 10 triggers per day over the full visible sky, the joint false-alarm rate will be
\begin{eqnarray}
R_{\mathrm{joint}} & = & \frac{1}{4} R_{\mathrm{GW}} \left(\frac{10}{86400 \punits{s}}\right) (33.6 \punits{s}) \cdot 0.04 \\
  & = & (4 \times 10^{-5}) \, R_{\mathrm{GW}} \,.
\end{eqnarray}
This means that the joint search can achieve the same false-alarm rate as the GW-only search by using a $\rho$ threshold $2.2$ units lower, e.g.\ $9.8$ instead of $12$.  Events will be detectable within a volume $(12/9.8)^3=1.84$ times as great within the sector of sky viewed by the radio search.  If $f_\mathrm{sky}=1/4$ (using a single radio facility), and the GW-only search is used for the rest of the sky, the overall increase in detected event rate due to the joint search is about 21\%.

A GW-triggered radio search using beams still has access to the overhead sky ($f_\mathrm{sky}=0.25$), while the beams are deliberately formed to overlap the GW sky-map, so $f_c=1$.  However, the beam-based search is cleaner and has a lower rate of real unassociated transients.  To estimate the net false-alarm rate, we assume that the dispersed radio pulse search is approximately gaussian (after data selection to avoid RF interference) and that the rate of real transients is smaller than the rate of triggers from noise excursions.  As a function of the SNR threshold,
\begin{equation}
R_{\mathrm{radio}}(\mathrm{SNR}) = N_B S \cdot \erfc{ \frac{\mathrm{SNR}}{\sqrt{2}} } \,,
\label{eqn:radio_false-alarm_rate}
\end{equation}
where $N_B$ is the number of beams and $S \approx 10^8 \punits{hr}^{-1}$ is a nominal value for the effective rate of independent filter outputs in the search considering all pulse times and DMs, per beam \citep{CutchinPC}.
With four beams, the joint false-alarm rate is
\begin{equation}
R_{\mathrm{joint}}(\rho,\mathrm{SNR}) = \frac{1}{4} \,100^{(11 - \rho)} \erfc{ \frac{\mathrm{SNR}}{\sqrt{2}} } \pxp{\frac{4 \cdot 10^8 \cdot 33.6}{3600}} \punits{yr}^{-1}.
\label{eqn:joint_false-alarm_rate_calc}
\end{equation}
For example, a threshold of $\mathrm{SNR}=7$ yields a radio trigger rate of $\usim 10^{-3}$ per hour and $R_{\mathrm{joint}} = (2.4 \times 10^{-6}) \, R_{\mathrm{GW}}$.  In this case, a joint false-alarm rate of one per hundred years can be achieved with $\rho\approx 9.2$.  Events will be detectable within a volume $(12/9.2)^3=2.22$ times as great within the sector of sky visible to the radio array, but only if one of the beams covers the true position of the source.  If the beams collectively contain a fraction $C$ of the GW sky-map probability, then the joint search will increase the total number of events detected by $(f_\mathrm{sky} C \cdot 122)$\%.  With four LWA1 beams, $C$ may typically be about $0.8$, yielding an increase of $\approx 24$\%.  This result is similar to the wide-area search above, even though trigger rates have been modeled differently.

\begin{figure}[htp]
\begin{center}
\includegraphics[width=0.6\textwidth]{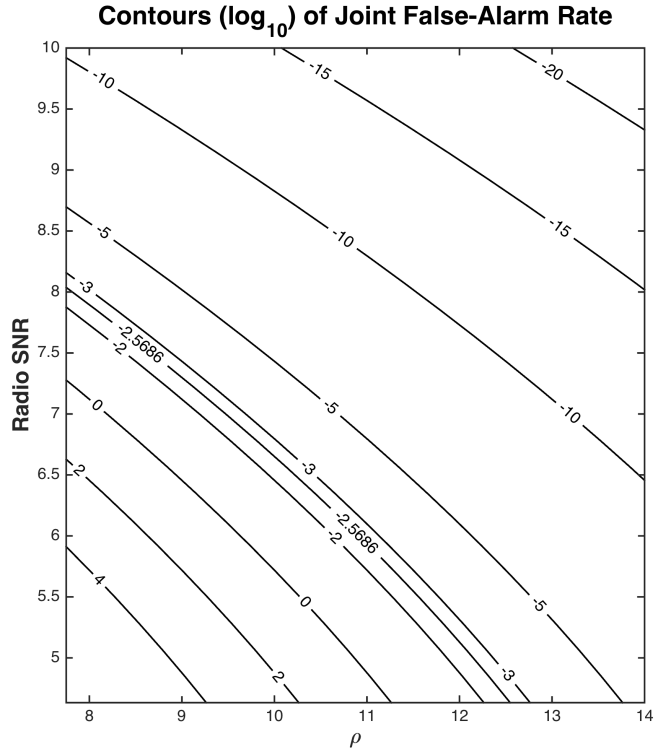}
\end{center}
\caption{\label{fig:effective_false_alarm}
Contour curves $\pxp{\log_{10}}$ of joint false-alarm rates for the case of a GW-Triggered joint search between LIGO/Virgo and LWA1, \eqnref{eqn:joint_false-alarm_rate_calc}.  Using these curves, one can choose a joint false-alarm rate and then adjust individual GW and radio detection thresholds to optimize the efficiency for detecting joint signals and work around instrument constraints.  Of note is the contour at $-2.5686$, corresponding to a joint false-alarm rate threshold $\Lambda \sim \scinum{2.7}{-3} \punits{yr}^{-1}$.  This false-alarm rate threshold is obtained from matching Poisson statistics for a 1 year observation period against a chosen candidacy requirement of a $3-\sigma$ or better excursion above the noise background.}
\end{figure}

\Figref{fig:effective_false_alarm} shows contour curves for other combinations of LIGO/Virgo and LWA1 trigger thresholds yielding a desired false-alarm rate.  Alternatively, it is possible to define a joint detection statistic in more sophisticated ways to select different regions of the $(\rho,\mathrm{SNR})$ plane, but that is most useful when a specific model of joint emission is known or assumed for optimization purposes.

For an untriggered, beamed-radio joint survey, a radio trigger is irrelevant if it does not overlap the GW sky-map, so the coincidence search is effectively constrained to the area of the beams; with four LWA1 beams, $f_\mathrm{sky}\approx 0.0025$.  $f_c=1$ since any GW trigger in that sky region will be coincident.  Following the calculation in the previous paragraph, the coincident false-alarm rate is a factor of 100 smaller, but the chance of detecting a signal is reduced by the same factor of 100 due to the comparatively small sky area viewed by the fixed beams.  A lower threshold on $\rho$ can be used, but it is still much less likely for this search strategy to successfully detect a joint signal.

\section{SUMMARY}
\label{sec:summary}

This paper has discussed the prospect of performing multi-messenger astronomy of high-energy astrophysical transients using gravitational waves and radio transients.  We reviewed a variety of mechanisms that could lead to coincident emission of both GW and radio frequency transients from select sources (i.e.\ NS-NS mergers and superconducting cosmic string cusp events).  Of these, we found that for compact object mergers \citet{MSPshirkov2010} describe the most promising model for radio transient emission observable with the current generation of instruments.

While EM counterparts to GW triggers are being sought in a wide range of wavelengths, we highlight that low-frequency radio instruments provide two interesting capabilities.  First, the expected dispersion delay of an extragalactic, MHz radio source is at least 10 minutes, and may be as long as an hour or more.  This leads to the real possibility of pointing a synthesized beam at the reconstructed GW source location \textit{before} the EM pulse arrives, and so observe any prompt emission in this band.  The other unique opportunity presented by radio dipole arrays is the large effective field-of-view that can be brought to bear, either with signal processing to generate sky images or by surveying with relatively large synthesized beams.

We considered three possible strategies for utilizing flexible low-frequency dipole array facilities to find radio transient counterparts to GW signals: a joint ``all-sky'' survey (above the horizon), a radio observation response triggered by a GW alert, and a joint survey with beamed radio observations.  Each strategy has advantages and disadvantages, depending on source characteristics and instrument capabilities.  In all cases, the conjoining of radio observations with GW observations has the effect of reducing the threshold on the GW detection statistic, increasing the sensitivity volume for LIGO/Virgo
compared to a GW-only search.  This increased range, along with the exciting possibility of observing prompt emission from a NS-NS merger in the nearby universe, provides strong motivation for carrying out this unique search.

Gravitational waves will be observed by LIGO, Virgo and future GW detectors with or without electromagnetic counterparts.  The GW signatures alone will enable tests of general relativity and give a crude picture of the population of sources.  The presence or absence of detectable prompt radio transients can then
test the various radio emission models for these systems.  This would allow studying the evolution of the orbital kinematics of masses within the binary and interactions with the magnetohydrodynamic environment enveloping the binary system.  


\acknowledgements

We would like to thank Sean Cutchin for insightful comments and shared knowledge. We would like to acknowledge the funding support provided by the U.S.\ National Science Foundation through grants PHY-1068549 and PHY-1404121, as well as Cooperative Agreement PHY-0757058.

\bibliographystyle{apj}
\bibliography{references.bib}

\appendix
\numberwithin{equation}{section}
\section{The Sensitivity of LWA1 and LOFAR to Radio Transients}
\label{sec:sensitivity_LWA}

At low frequencies, Galactic noise is the dominant contribution to system noise.  \cite{EllingsonSensitivity} established a system model and procedure for estimating the system equivalent flux density
(SEFD), the 1$\sigma$ ``bottom line'' flux density, which accounts for the combined effects of all noise sources.  Ellingson uses a spatially uniform sky brightness temperature $T_b$ in his model, dependent on observing frequency $\nu$, where
\begin{equation}
T_b = 9751\ {\rm K}\ \left(  \frac{\nu}{38\ {\rm MHz}}   \right)^{-2.55}
\end{equation}
and ignores the ground temperature contribution as negligible.  The receiver noise is about 250 K, but has little influence on the SEFD.  The method, when applied to LWA1, shows the correlation of Galactic noise between antennas significantly desensitizes the array for beam-pointings that are not close to the zenith.  It is also shown that considerable improvement is possible using beam-forming coefficients that are designed to optimize signal-to-noise ratio under these conditions.  The result implies, for beams near the zenith, the flux density necessary to produce a specific signal-to-noise ratio (SNR) is approximately
\begin{equation}
f_\nu \approx 7\ {\rm Jy}\  \left( \frac{\rm SNR}{10} \right) B_{20}^{-1/2} \Delta t^{-1/2}
\label{limitingflux1}
\end{equation}
at either 38~MHz or 74~MHz, for a bandwidth $B_{20}$ in units of 20~MHz, and an integration time $\Delta t$ in seconds.   The result is slightly dependent on whether one uses beam-forming coefficients entirely designed to remove delays or optimized to produce the best SNR, but the above equation is sufficient for current purposes.  It is expected that $f_\nu$ is 10 times larger than eq.~(\ref{limitingflux1}) at zenith angle $\theta=50^\circ$ for 38~MHz, and $\theta=60^\circ$ for 74~MHz.

LOFAR is similarly noise dominated by the Galaxy, where the temperature is \citep{LOFAR.Memo113}
\begin{equation}
T_{sky} = T_{s0}\ \lambda_{m}^{2.55}
\end{equation}
and $T_{s0}=60 \pm 20$~K, for $\lambda_{m}$ in meters.  Observing with 13 core and 7 remote stations gives the flux density to be
\begin{equation}
f_\nu \approx 2\ {\rm Jy}\ \left( \frac{\rm SNR}{10} \right) B_{4}^{-1/2} \Delta t^{-1/2}
\label{limitingfluxLOFAR}
\end{equation}
at 120~MHz, for a bandwidth $B_{4}$ in units of 4~MHz, and an integration time $\Delta t$ in seconds.

In observing a radio transient the best SNR is obtained when the integration time $\Delta t$ is matched to the transient pulse duration, or width.  In practice, a search within the data uses a range of trial widths to find the appropriate integration width.  Predicting the expected SNR for a specific radio transient model requires knowing the expected flux density and arriving pulse width, as shown explicitly in eq.~(\ref{limitingflux1}).  The arriving pulse width depends on the emitted width, and the width broadening effects due to dispersion and scattering.

The temporal pulse broadening of an emitted pulse depends on a combination of effects as described by \cite{JMCordes2003}
\begin{equation}
\Delta t = [\Delta t_{\rm intrinsic}^2 + \Delta t_{\rm DM}^{2} + \Delta t_{\delta \rm DM}^{2} + \Delta t_{\Delta \nu}^{2} + \tau_{\rm d}^{2}]^{1/2}
\label{temporal broadening}
\end{equation}
where $\Delta t_{\rm intrinsic}$ is the emitted pulse width, $\Delta t_{\rm DM}$ is dispersion smearing, $\Delta t_{\delta \rm DM}$ is dedispersion error, $\Delta t_{\Delta \nu}$ is the receiver filter response time, and $\tau_{\rm d}$ is the scatter-broadening term.  The effects due to dedispersion error and receiver filter response time are negligible and are excluded from further calculations.

The scatter-broadening model of Cordes and McLaughlin describes pulses from extragalactic sources
\begin{equation}
\frac{\tau_{d \rm xgal}}{\tau_{d \rm Gal}} \approx 3.7 \left( 1 + \frac{SM_{\rm xgal} D_{\rm xgal}}{SM_{\rm Gal} D_{\rm g}} \right)^{6/5} \left( \frac{D_{\rm g}}{D} \right)^{1/5}
\label{CM xgal scatter broadening}
\end{equation}
where $\tau_{d \rm Gal}$ is approximated by the empirical fit \citep{DRLorimer2013}
\begin{equation}
\log \tau_{d\rm Gal} = -6.5 + 0.15 \log {\rm DM} + 1.1\,(\log {\rm DM})^2 - 3.9\, \log \nu_{\rm GHz} \punits{ms} \,.
\label{CM Gal scatter broadening}
\end{equation}
Here, $\rm DM$ is the dispersion measure of the signal in ${\rm pc}/{\rm cm}^3$ units, $D_{\rm g}$ is the distance the signal travels through our Galaxy, $D_{\rm xgal}$ is the distance the signal travels through the host galaxy, and $D$ is the distance to the source.  In their model, the intergalactic medium insignificantly contributes to scattering, and thin screens are placed within the host galaxy and the Milky Way.  For simplicity, it is assumed most sources will be approximately perpendicular to the disk of the Milky Way; thus, $D_{\rm g} \sim 1$~kpc and ${\rm DM} = 30$.  It can also be argued that the scattering measures and pulse travel distances are roughly equal for the host galaxy and the Milky Way, reducing eq.~(\ref{CM xgal scatter broadening}) to
\begin{equation}
\tau_{d \rm xgal} \approx \scinum{2}{-2}\, D_{\rm Gpc}^{-1/5} \nu_{40}^{-3.9} \\  \ {\rm s},
\label{reduced CM}
\end{equation}
where $D_{\rm Gpc}$ is the distance in gigaparsecs and $\nu_{40}$ is in units of 40~MHz.

\cite{DRLorimer2013} suggests that at greater extragalactic distances the intergalactic medium dominates as the scattering medium, and therefore uses a thin screen model with the thin screen placed halfway between the host galaxy and the Milky Way
\begin{equation}
\rm log \tau_{d\rm Gal} = -9.5 + 0.15 \log {\rm DM} + 1.1\, (\log {\rm DM})^2 - 3.9 \rm log \nu_{\rm GHz} \punits{ms} \,.
\label{Lorimer scatter broadening}
\end{equation}
When the two scattering models are added in quadrature, as in \eqnref{temporal broadening}, it can be seen that the Cordes and McLaughlin scattering model dominates for distances out to 0.4~Gpc, after which the Lorimer scattering model dominates.  Since the coincidence search described in this paper is limited by Advanced LIGO at 0.2~Gpc, the contributions of the Lorimer model can be omitted for simplicity when analyzing coincident signals.  However, the detection distance values for LWA1 and LOFAR provided in the Sources section consider both scattering models since they are expected to see farther, in most cases, than Advanced LIGO.

Dispersion smearing follows the well-known relationship
\begin{equation}
\Delta t_{\rm DM} = 8.3 \,{\rm DM}\, \Delta \nu_{\rm MHz} \,\nu_{\rm GHz}^{-3} \\ \ \mu{\rm s}
\label{dispersion smearing}
\end{equation}
where $\Delta \nu$ is the width of a frequency channel; $\Delta\nu = 4.9\ {\rm kHz}$ for LWA1 \citep{EllingsonLWA1}, $\Delta\nu = 0.76\ {\rm kHz}$ for LOFAR \citep{LOFAR.Memo113}.  Dispersion smearing is the sum of contributions from the Milky Way, the host galaxy, the intergalactic medium, and any galaxies along the line of sight.  Following the assumptions discussed above, it is assumed the dispersion measure in the host galaxy is $\sim 30 \ \rm pc \ \rm cm^{-3}$, similar to the Milky Way.  The contribution from the intergalactic medium assumes all the baryons in the universe form a uniformly distributed, completely ionized gas throughout intergalactic space.  Then, the free electron number density at low $z$ is $3 H_o^2 \Omega_b / 8 \pi G m_p \approx 2 \times 10^{-7} \ \rm cm^{-3}$ \citep{Ioka2003, SInoue2004}.  Thus, for a line of sight of length $D$,
\begin{equation}
{\rm DM}_{\rm IGM} = 20 \left( \frac{D}{100\,\rm Mpc} \right) {\rm pc cm^{-3}}.
\label{DM IGM}
\end{equation}
It is unlikely there are any other galaxies in the line of sight to the host galaxy; thus, no dispersion from other galaxies along the line of sight is assumed.  Therefore, the total dispersion-measure is
\begin{equation}
{\rm DM} = 60 + 200\, D_{\rm Gpc} \\ \ {\rm pc \ cm}^{-3}.
\label{total DM}
\end{equation}
The contribution from dispersion smearing to the total temporal pulse broadening for LWA1 is
\begin{equation}
\Delta t_{\rm DM} = [0.038 + 0.13\, D_{\rm Gpc}] \nu_{40}^{-3} \\ \ {\rm s},
\label{DM broadening}
\end{equation}
and for LOFAR, it is
\begin{equation}
\Delta t_{\rm DM} = [0.0059 + 0.02\, D_{\rm Gpc}] \nu_{40}^{-3} \\ \ {\rm s}.
\end{equation}
The total expression for the temporal broadened pulse width as measured by LWA1 is given by
\begin{equation}
\Delta t = \left[ \Delta t_{\rm intrinsic}^2 + \left([0.038 + 0.13\, D_{\rm Gpc}] \nu_{40}^{-3} \right)^{2} + \left(2 \cdot 10^{-2}\, D_{\rm Gpc}^{-1/5} \nu_{40}^{-3.9} \right)^{2} \right]^{1/2} \\ \ {\rm s},
\label{final tb}
\end{equation}
and the pulse width as measured by LOFAR is given by
\begin{equation}
\Delta t = \left[ \Delta t_{\rm intrinsic}^2 + \left([0.0059 + 0.02\, D_{\rm Gpc}] \nu_{40}^{-3} \right)^{2} + \left(2 \cdot 10^{-2}\, D_{\rm Gpc}^{-1/5} \nu_{40}^{-3.9} \right)^{2} \right]^{1/2} \\ \ {\rm s}.
\end{equation}


\end{document}